# Microstructural studies on lattice imperfections in irradiated Titanium and Ti-5%Ta-2%Nb by X-Ray Diffraction Line Profile Analysis

P.Mukherjee*, A. Sarkar*, P.Barat*, Baldev Raj[#] and U.Kamachi Mudali[#]

*Variable Energy Cyclotron Centre, 1/AF Bidhannagar, Kolkata- 700 064, India.

#Metallurgy and Materials Group, Indira Gandhi Centre for Atomic Research, Kalpakkam-603 102

## ABSTRACT

The microstructural parameters like the average domain size, effective domain size at a particular crystallographic direction and microstrain within the domains of titanium and Ti-5%Ta-2%Nb, irradiated with 116 MeV $O^{5+}$ ion, have been characterized as a function of dose by X-Ray Diffraction Line Profile Analysis using different model based approaches. Dislocation Density and stacking fault probabilities have also been estimated from the analysis. The analysis revealed that there was a significant decrease of the average domain size with dose as compared to the unirradiated sample. The estimated values of dislocation density increased significantly for the irradiated samples and was found to be an order of magnitude more as compared to the unirradiated one. However, the dislocation density saturated with increase in dose. The deformation (stacking) fault probabilities were found to be negligible even with the increase in dose of irradiation.

## I. INTRODUCTION

Materials irradiated with the energetic particles undergo changes in the structure and the properties. Energetic particles transfer energy to the materials primarily by the process of ionization, electronic excitation and also by the displacements of atoms from their original sites. These cause a change in the internal microstructure, phase distributions, dimensions, mechanical and corrosion properties [1-4] of the target materials. The nature of radiation damage in the materials is affected by the type of ions used for irradiation, alloying elements and the impurity variations [5]. In case of the light ions such as protons, the damage profiles are homogeneous. On the other hand, heavy ions having energy of the order of few MeV produce displacement cascades consisting of highly localized interstitials and vacancies [6].

Titanium has been selected and used for the construction of the electrolytic dissolver of the spent fuel reprocessing plant for the fuel of the Fast Breeder Test Reactor (FBTR) at Kalpakkam, India[7]. Titanium shows excellent corrosion resistance in various concentrations of nitric acid, particularly at the boiling condition in which spent fuel is dissolved [8]. However, it undergoes corrosion attack in vapour phase and condensate phase of the acid due to the formation of loosely adherent oxide film on the surface [7,9]. The presence of excess iron (>0.05%) greater than the solubility of iron in titanium also leads to severe corrosion, due to the formation of iron-rich intermetallics that can dissolve preferably in boiling nitric acid [9]. In order to overcome these specific corrosion problems, new alloys are being developed which can resist corrosion under such conditions. Ti-5%Ta-2%Nb is a new generation alloy developed to overcome the vapour and condensate phase corrosion by the stabilization of oxide film by Ta and Nb addition, and also by increasing the solubility of iron in a near-alpha microstructure[9]. The electrolytic dissolver vessel made of Ti and Ti-5%Ta-2%Nb would be used in severe corrosive nitric acid in a highly radioactive condition of $10^5$-$10^6$ Rad/hour. Radiation induced defects would influence the migration of carriers such as electrons or ions through the passive films on the surface of the materials. The reliability of the material of construction at such zones have been reported to be affected by the movement of electrons or ions destabilizing the protective passive film[10,11]. As an example, an alpha particle can penetrate the passive $TiO_2$ film and produce defects. Several other possibilities leading to the destabilization of the protective passive film has been discussed by Elfenthal et al.[12]. Ion implantation and laser illumination are two methods attempted in the past[11] for the creation of the irradiation induced defects.

These irradiation induced defects also cause a drastic change in the microstructure of the materials Ti and Ti-5%Ta-2%Nb, used for fast reactor fuel reprocessing applications. Therefore, it is interesting to study the microstructure of these irradiated materials. Subsequently, the corrosion studies will be carried out on the surfaces of the irradiated specimen dipped in $HNO_3$ and also on artificially generated protective film of the specimens of the same alloys.

In the present study, we have carried out irradiation with 116 MeV $O^{5+}$ on Ti and Ti-5%Ta-2%Nb at different doses and the microstructural parameters have been characterised by X-ray Diffraction Line Profile Analysis (XRDLPA). XRDLPA is a powerful technique to evaluate the average microstructural parameters in a statistical manner. Different techniques of XRDLPA have been widely applied for the evaluation of the microstructural parameters in different deformed metals and alloy systems [13-14]. In our earlier studies, we have characterized the microstructure of the proton-irradiated and oxygen-irradiated Zr-1%Nb-1%Sn-0.1%Fe[15-16].

In this work, we have characterised the microstructural parameters by XRDLPA using different model based approaches like Williamson-Hall Technique, Modified

Rietveld Method and Double Voigt Method on irradiated Ti and Ti-5%Ta-2%Nb. The domain size, microstrain, dislocation density and the stacking fault probabilities of these irradiated alloys have been estimated as a function of dose. The damage profile as a function of depth from the surface has been characterized in terms of displacements per atom (dpa) at different doses.

## II. EXPERIMENTAL

Commercially pure titanium specimens of size 10mm x 10mm were cut from a sheet of 3mm thickness and were used for the present investigation. Ti-5%Ta-2%Nb was produced at Nuclear Fuel Complex, Hyderabad, India and disc specimens (2mm long, 28 mm diameter) were cut from an extruded rod (chemical composition is given in Table 1).

The samples of titanium and Ti-5%Ta-2%Nb were mounted on an aluminium flange and then irradiated with 116 MeV $O^{5+}$ ions from Variable Energy Cyclotron (VEC), Kolkata, India. The irradiation doses were $1 \times 10^{17}$, $1 \times 10^{18}$ and $1 \times 10^{19}$ $O^{+5}$ ions/m$^2$. The ion current used in the experiment was 150 nA. The flange used for the irradiation was cooled by a continuous flow of water. During irradiation, the temperature of the sample did not rise above 313K as measured by the thermocouple connected very close to the sample. The range of the ions in these materials and the displacement per atoms (dpa) were obtained by Monte-Carlo simulation technique using the code TRIM 95[17].

X-Ray Diffraction (XRD) profile for each irradiated sample has been recorded by PHILIPS 1710 diffractometer using CuK$_\alpha$ radiation. The range of 2θ was from 25° to 100° and a step scan of 0.02° was used. The time per step was 4 seconds.

## III. METHOD OF ANALYSIS

The diffraction of the radiation from the matter corresponds to a Fourier transform from the real space to the momentum space; hence the XRD pattern of a sample represents a complete mapping of its crystal structure and the microstructure in the momentum space. In most investigations, the information of the crystal structure is extracted from the diffraction pattern namely the angular positions and the intensities of the Bragg peaks. In the present study, we are interested in the microstructure. Generally, the broadening of a Bragg peak arises due to the instrumental broadening, broadening due to the small domain size and the microstrain within the domain. However, a detailed information is extractable from the line shapes of the Bragg peaks. The analysis of the line shapes allows one to characterise the microstructure more comprehensively in terms of the mean square microstrain and the average domain size. Williamson-Hall Technique, Modified Rietveld Method using whole powder pattern fitting technique and Double Voigt Method have been adopted in the present study in order to analyse the diffraction data of titanium and Ti-5%Ta-2%Nb at different doses of irradiation. All these techniques are based on the analysis of the shapes of the broadened diffraction profiles. The instrumental broadening correction was made using a standard defect free Si sample.

*Williamson-Hall Technique*

Williamson and Hall [18] assumed that both size and strain broadened profiles are Lorentzian. Based on this assumption, a mathematical relation was established between the integral breadth ($\beta$), volume weighted average domain size ($D_v$) and the microstrain ($\varepsilon$) as follows.

$$\frac{\beta \cos\theta}{\lambda} = \frac{1}{D_v} + 2\varepsilon\left(\frac{2\sin\theta}{\lambda}\right) \qquad (1)$$

The plot of $\left(\frac{\beta\cos\theta}{\lambda}\right)$ versus $S = \left(\frac{2\sin\theta}{\lambda}\right)$ gives the value of the microstrain from the slope and domain size from the ordinate intercept.

*Modified Rietveld Method*

In this method, the diffraction profile has been modelled by a pseudo-Voigt (*pV*) function using the program LS1[19].

This program includes the simultaneous refinement of the crystal structure and the microstructural parameters like the domain size and the microstrain within the domain. The method involves the Fourier analysis of the broadened peaks. Considering an isotropic model, the lattice parameters (a and c), surface weighted average domain size ($D_s$) and the average microstrain $\langle\varepsilon_L^2\rangle^{\frac{1}{2}}$ were used simultaneously as the fitting parameters to obtain the best fit. The effective domain size ($D_e$) with respect to each of the fault-affected crystallographic plane was then refined to obtain the best fitting parameter. The preferred orientation produces a systematic distortion of the reflection intensities. The preferred orientation correction parameter $P(\alpha)$ [20-21] has the form:

$$P(\alpha) = \left(r^2 \cos^2\alpha + \frac{\sin^2\alpha}{r}\right)^{-3/2} \qquad (2)$$

where $\alpha$ is the angle between (hkl) plane and the preferred oriented plane and *r* is an adjustable parameter. XRD peak profiles of titanium and Ti-5%Ta-2%Nb show a strong crystallographic texture along certain crystallographic directions particularly (002), (101), (102) and (103) planes. The h,k,l values of these planes were incorporated in the program as the preferred oriented planes and the best fit was sought in each case.

The average dislocation density (ρ) has been estimated from the relation [22]

$\rho = (\rho_D \rho_S)^{\frac{1}{2}}$, where, $\rho_D = \frac{3}{D_s^2}$ (dislocation density due to domain) and $\rho_S = k\langle \varepsilon_L^2 \rangle / b^2$ (dislocation density due to strain), k is the material constant and $b$ is the modulus of the Burger's vector, $\frac{1}{3}[11\bar{2}0]$. Similarly, $\rho_e$, the dislocation density at each crystallographic plane has been estimated. The values of dislocation density in all these cases estimated from the analysis are much approximate values as we have considered the random distributions of the dislocations.

The effective domain size $D_e$ is related to the surface weighted average domain size $D_s$ and the stacking faults (deformation fault $\alpha$ and growth fault $\beta$) by the following relations [23]:

$$\frac{1}{D_e} = \frac{1}{D_s} + [L_0|d(3\alpha + 3\beta)/C^2] \text{ for } L_0 \text{ even} \qquad (3)$$

$$\frac{1}{D_e} = \frac{1}{D_s} + [L_0|d(3\alpha + \beta)/C^2] \text{ for } L_0 \text{ odd} \qquad (4)$$

where $d$ is the lattice spacing, $C$ is the lattice constant and $L_0$=h+k+l.

The deformation fault $\alpha$ and the growth fault $\beta$ were then separated by the least square analysis considering the fault affected reflections. Since growth fault is absent in a h.c.p alloy[24-27], the values of deformation fault $\alpha$ are reported.

*Double Voigt Method*

A Cauchy or a Gaussian function exclusively cannot model the peak broadening. Therefore, the size and the strain effects are approximated by a Voigt function [28], which is basically a convolution of Gaussian and Cauchy function. The equivalent analytical expressions for Warren-Averbach size-strain separation [29] are then obtained. The Fourier coefficients $F(L)$ in terms of a distance, $L$, perpendicular to the diffracting planes are obtained by Fourier transform of the Voigt function[28] and can be written as

$$F(L) = \exp(-2L\beta_C - \pi L^2 \beta_G^2) \qquad (5)$$

where, $\beta_C$ and $\beta_G$ are the Cauchy and the Gauss components of total integral breadth $\beta$ respectively.

$\beta_C$ and $\beta_G$ can be written as:

$$\beta_C = \beta_{SC} + \beta_{DC} \qquad (6)$$

$$\beta_G^2 = \beta_{SG}^2 + \beta_{DG}^2 \qquad (7)$$

where, $\beta_{SC}$ and $\beta_{DC}$ are the Cauchy components of size and the strain integral breadth respectively and $\beta_{SG}$ and $\beta_{DG}$ are the corresponding Gaussian components.

The size and the distortion coefficients are obtained considering at least two reflections from the same family of crystallographic planes. The surface weighted average domain size $D_S$ and the microstrain $\langle \varepsilon_L^2 \rangle^{\frac{1}{2}}$ are given by the equations:

$$D_S = 1/2\beta_{SC} \qquad (8)$$

$$\langle \varepsilon_L^2 \rangle = [\beta_{DG}^2/(2\pi) + \beta_{DC}/(\pi^2 L)]/S^2 \text{ where } S = \frac{2\sin\theta}{\lambda} \qquad (9)$$

The volume weighted domain size [30] is given by:

$$D_V = \frac{1}{\beta_S} \text{ where } \beta_S = \frac{\beta \cos\theta}{\lambda}, \text{ integral breadth in the units of } S, (\text{Å})^{-1}.$$

The volume weighted column-length distribution functions are given by:

$$P_v(L) \propto L \frac{d^2 A_S(L)}{dL^2} \qquad (10)$$

For a size-broadened profile, the size coefficient is given as:

$$A_S(L) = \exp(-2L\beta_{SC} - \pi L^2 \beta_{SG}^2) \qquad (11)$$

From equation (11), we get,

$$\frac{d^2 A_S(L)}{dL^2} = [(2\pi L \beta_{SG}^2 + 2\beta_{SC})^2 - 2\pi \beta_{SG}^2] A_S(L) \qquad (12)$$

Selivanov and Smislov [31] showed that equation (12) is a satisfactory approximation of size distribution functions.

## IV. RESULTS AND DISCUSSION

The range of 116 MeV $O^{5+}$ ion in titanium and Ti-5%Ta-2%Nb (obtained by TRIM 95 calculation) was found to be around $78\mu m$. The radiation damage has been assayed by the damage energy deposited causing displacements of atoms. The total target displacements of the collision events calculated by the programme TRIM 95, is shown in Fig.1. The damage is measured by the number of displacements per atom (dpa). The average dpa for the highest dose sample in titanium was found to be $3.9 \times 10^{-3}$. We have calculated the dpa averaged over the total range of $78\mu m$.

Fig.2 represents typical X-ray diffraction profiles of unirradiated Ti and irradiated Ti at a dose of $1\times10^{19}O^{5+}/m^2$ (highest dose). There is a clear broadening of the diffraction peaks of the irradiated sample as compared to the unirradiated one. The findings obtained by different XRDLPA techniques are illustrated below.

*Williamson-Hall technique*

Fig.3 and Fig.4 show WH plots for both unirradiated and irradiated Ti and Ti-5%Ta-2%Nb at different doses respectively. For most of the cases, it is seen that $\frac{\beta\cos\theta}{\lambda}$ vs $S$ shows a linear dependence. This implies that the shape of the domains remained isotropic even after irradiation. It is further observed that the slope of the line connecting two orders of (00l) type reflections (i.e. <002> & <004>) increased with irradiation for Ti but did not change significantly for Ti-5%Ta-2%Nb alloy. This indicates a lattice distortion along <00l> of Ti after irradiation. A strong lattice distortion may be similarly predicted along <101>. The average values of $D_v$ and $\varepsilon$ obtained from the intercept and the slope of WH plots are shown in Table-2. It is observed that for both pure titanium and Ti-5%Ta-2%Nb, $D_v$ decreased with dose. The values of $\varepsilon$ were found to increase slightly for pure titanium with dose but there was a decreasing trend of $\varepsilon$ for Ti-5%Ta-2%Nb at higher doses.

*Modified Rietveld Method*

We have carried out analysis on XRD patterns of unirradiated and irradiated Ti and Ti-5%Ta-2%Nb with the help of modified Rietveld method using the program LS1 [19]. Fig. 5 shows a fitted diffraction profile of Ti-5%Ta-2%Nb at the highest dose of irradiation ($1\times10^{19}O^{5+}/m^2$). The values of the weighted pattern ($R_{wp}$ % and $R_e$ %) of this figure after refinement were found to be 10.89 and 8.46 respectively and the goodness of fit was 1.28 for this fitted profile. The residuals are also shown below, in the Fig. 5. The variations of $D_s$ and $\rho$ for these samples have been plotted as a function of dose in Fig.6 and Fig.7 respectively.

Significant changes were found in the values of $D_s$ and $\rho$ with dose in oxygen-irradiated samples as compared to the unirradiated one for both the materials. There was a drastic decrease in domain size at a dose of $1 \times 10^{17} O^{5+}/m^2$ but the values saturated with increasing dose of irradiation. The dislocation density increased significantly for the irradiated samples and the increase was found to be almost an order of magnitude more in case of the irradiated samples as compared to the unirradiated one. These values were also found to saturate with dose. The reasons of the above findings can be explained as follows.

The range of 116 MeV oxygen ion in pure titanium is $78 \mu m$. Oxygen being a heavy ion, transfers sufficient energy to the primary knock on atoms which in turn produce displacement cascades, consisting of highly localized interstitials and vacancies. As the primary knock on proceeds through the sample, loosing energy in successive collisions, the displacement cross-section increases[32]. Thus the distance between successive displacements decreases and at the end of the track, the recoil collides with practically every atom in its path, creating a very high localised concentration of vacancies and interstitials. Moreover, the energy transferred to the lattice atoms is much larger at the end of the trajectory of the projectile. A concentration gradient of defects in the sample was thus created within a small reaction path of $78 \mu m$, which helped in migration of defects by radiation enhanced diffusion process and agglomeration of them.

SIZZMANN [33] reported that in the radiation enhanced diffusion process, excess point defect concentrations are produced by irradiation with the high energy particles. This not only causes an enhancement of the diffusion process but also opens up new channels by the creation of different defect species, which are not available in normal activated diffusion. Thus, the diffusion coefficient of a particular lattice atom gets enhanced by the linear superposition of various conceivable diffusion channels [33] due to the presence of irradiation induced vacancies, di-vacancies, interstitials etc. Thus the enhancement of the migration of vacancies, caused the nucleation of vacancy clusters, which collapses in the shape of dislocation loops[34]. As a result, we found a significant increase in dislocation density in the irradiated samples as compared to the unirradiated one. However, the dislocation density almost saturated with the increase in the dose of irradiation (Fig.7).

In the irradiated sample, the mechanism responsible for the generation of dislocations is solely dependent on the collapse of agglomerated vacancies, as Frank Reed source mechanism for the multiplication of dislocations is absent due to the non-availability of any stress field. The generation of dislocations by the collapsing of vacancy clusters is only possible, when there is an excess vacancy concentrations than the equilibrium values, as in the irradiated sample. Hence, we could observe an order of magnitude increase in the dislocation density even at a dose of $1 \times 10^{17}$ $O^{5+}/m^2$. During irradiation, two competing processes occur simultaneously, one is the generation of vacancies, agglomeration of vacancies and then collapsing into dislocation loops and the other is, their annihilation at the possible sinks. Initially, at the low dose of irradiation ($1 \times 10^{17}$ $O^{5+}/m^2$), the rate of generation of dislocation loops dominates over the rate of annihilation of the point defects as the sink density is low. So, we found an increase in the dislocation density. With increasing dose of irradiation, though more vacancies are created, annihilation rate of vacancies also increases as the sink density increases with irradiation. Hence, saturation was observed in the dislocation density with the increase in the dose of irradiation.

The domains in the irradiated samples formed due to the entanglement of the dislocations present in them. The size of the domains in the irradiated samples decreased due to the interaction of dislocation loops with the dislocation substructure present in the sample before irradiation. The decrease was quite drastic at lower doses and almost saturated at higher doses, as the generation of dislocation did not vary significantly with the increase in dose.

The effective domain size $D_e$, along the different crystallographic directions was found to decrease with dose as compared to unirradiated material but the shape of the domains were almost isotropic for both these alloys. We have plotted the projections of $D_e$ (along different directions) on the plane containing the directions <002> and <100>. Only the projections in the first quadrant are shown in Fig.8 and Fig.9 respectively for Ti and Ti-5%Ta-2%Nb at different doses. It was clearly observed that, $D_e$ was almost isotropic (spherical) with the values $<D_e>_{002} \cong 731$Å and $<D_e>_{100} \cong 607$Å for unirradiated titanium, $<D_e>_{002} \cong 299$ Å and $<D_e>_{100} \cong 314$ Å at a dose of $1 \times 10^{17}$ $O^{5+}/m^2$, $<D_e>_{002} \cong 301$Å and $<D_e>_{100} \cong 301$Å at a dose of $1 \times 10^{18}$ $O^{5+}/m^2$ and $<D_e>_{002} = 290$ Å and $<D_e>_{100} = 268$ Å at a dose of $1 \times 10^{19}$ $O^{5+}/m^2$. The domains of Ti-5%Ta-2%Nb were also spherically isotropic for both unirradiated and irradiated materials i.e. $<D_e>_{002} \cong 642$Å and $<D_e>_{100} \cong 649$Å for unirradiated Ti-5%Ta-2%Nb, $<D_e>_{002} \cong 252$Å and $<D_e>_{100} \cong 252$Å at a dose of $1\times10^{17}$ $O^{5+}/m^2$ and $<D_e>_{002} \cong 200$Å and $<D_e>_{100} \cong 200$Å at a dose of $1\times10^{18}$ $O^{5+}/m^2$ and $<D_e>_{002} = 200$Å and $<D_e>_{100} = 176$Å at a dose of $1\times10^{19}$ $O^{5+}/m^2$. Thus, it was revealed that the shape of the domains did not change with the increase in dose though the variations in the size of the domains were significant with dose as compared to unirradiated sample.

The estimated values of the dislocation density at each crystallographic plane for both the materials as a function of dose are shown in Table-3.

The microstrain values at $L=50$Å along the different crystallographic directions for both the materials at different doses are shown in Table-3. The values showed an increasing trend for irradiated titanium as compared to unirradiated one. But the values were found to decrease with increasing dose of irradiation for Ti-5%Ta-2%Nb.

The aspects of the cell structures or domains of deformed materials depend on the various extrinsic (e.g., strain rate, temperature, crystal orientation) and intrinsic (crystal structure, stacking fault energy, chemical composition) parameters. However, their appearance seems to obey universal principles. This is reflected by the validity of an empirical law relating $D_e$ and the dislocation density $\rho_e$ by the relation:

$$D_e = K(\rho_e)^{-\frac{1}{2}} \tag{12}$$

where $K$ is a material constant.

A compilation of available data[35] for both f.c.c and high temperature deformed b.c.c metals gave a master curve of $D_e$ vs. $(\rho_e)^{-\frac{1}{2}}$, where $K$ equals to 20. Fig.10 shows such a curve for the irradiated Ti and its alloys, simultaneously fitted for both the materials (unirradiated and irradiated) from the values of $D_e$ and $(\rho_e)^{-\frac{1}{2}}$ (obtained from modified Rietveld method), which yields $K$ equal to 0.59.

The deformation fault ($\alpha$) was found to be negligibly smaller for both the materials at different doses as seen in Table-3. The faulting probability of these alloys did not change even with increasing dose of irradiation.

*Double Voigt Method*

The general conclusions obtained from the simple WH plot can be further substantiated by a detailed analysis. In this analysis, both the size and strain broadened profiles were approximated by a Voigt function and the Cauchy and the Gaussian components of the size and strain broadened profiles ($\beta_{SC}, \beta_{SG}, \beta_{DC}$ and $\beta_{DG}$) were separated along <001> and listed in Table-4. From Table-4, it is observed that, in general the size broadened profiles had both Cauchy and Gaussian components of the integral breadths, except for unirradiated Ti-5%Ta-2%Nb and the sample with a dose of $1 \times 10^{17} O^{5+}/m^2$. In these two cases, the size-broadened profiles are generally Cauchy in nature, indicating a broad size distribution of the domains [36]. The size distribution has narrowed down with increasing dose.

The volume weighted column-length distribution function $P_v(L)$ along <001> has been shown in Fig.11 for both these materials. In comparison to unirradiated and irradiated Ti-5%Ta-2%Nb, it is clear that the domain size distributions are much wider for unirradiated titanium and also for the samples with dose of $1 \times 10^{17} O^{5+}/m^2$ and $1 \times 10^{18} O^{5+}/m^2$, indicating non-uniform domain size distribution along ‚001‚. The size distribution was slightly narrowed down at the highest dose of irradiation for pure titanium. On the other hand, the size distribution along <001> of Ti-5%Ta-2%Nb for unirradiated and irradiated with a dose of $1\times10^{17}O^{5+}/m^2$ and $1\times10^{18}O^{5+}/m^2$ were almost identical and it has narrowed down significantly at the highest dose of irradiation.

Using different model based approaches of XRDLPA techniques, the microstructure of the irradiated Ti and its alloy have been characterized. All these techniques are based on the profile shape and the broadening of the diffraction peak. These techniques have limitations in characterising the small defects particularly small interstitial clusters which do not cause broadening of the peak but contribute to the background values close to the Bragg peak [37]. Scattering of X-rays from interstitial clusters [38] are diffuse scattering very close to the Bragg peak (Huang Scattering). Thus, the complete information of the microstructure of the irradiated samples can be obtained from the X-ray diffraction techniques by the combined studies of the diffraction pattern in the Bragg peak region (coherent scattering) and in the background region (diffuse scattering close to the Bragg peak). As in our case, the experiments were carried out at room temperature, the diffuse scattering near the Bragg peak region due to small interstitial clustering are superimposed by thermal diffusion scattering. Hence, the line profile analysis could characterise only those microstructurral parameters which are responsible for the broadening of the diffraction peaks.

## V. CONCLUSION

Microstructure of the unirradiated and irradiated titanium and Ti-5%Ta-2%Nb has been reliably assessed by the different techniques of XRDLPA using different model based approaches. The microstructural parameters like average and effective domain sizes and microstrain within the domains have been characterised as function of dose.

The dislocation density and the stacking fault probabilities have been estimated from these values. The analysis revealed that there was a significant decrease of surface weighted average domain size ($D_s$) with the increase in dose. The damage associated with the oxygen beam (being heavy ion) was quite extensive and produced a highly localised concentration of defects, particularly vacancies and interstitials. The agglomeration of vacancies caused the nucleation vacancy clusters, which collapsed in the shape of dislocation loops and the dislocation density increased accordingly. However, the dislocation density saturated with dose of irradiation. The average dislocation density in most of the planes was found to be of the order of $10^{15} m^{-2}$, which was almost one order magnitude higher than the unirradiated sample. The deformation (stacking) fault probability was found to be negligible for both the materials even with the increasing dose of irradiation. The domain size-distribution was found to be narrower at the highest dose of irradiation for these materials. An empirical relationship has been established between $D_e$ with $(\rho_e)^{-\frac{1}{2}}$ for these systems.

**Table-1: Chemical composition of Ti-5%Ta-2%Nb alloy**

| Element | Ta | Nb | Fe | O | N(ppm) | C(ppm) | H(ppm) | Ti |
|---|---|---|---|---|---|---|---|---|
| Content (wt %) | 4.4 | 1.94 | 0.03 | 0.05 | 50 | 125 | 10 | balance |

## Table-2 : Results of Williamson-Hall Plot

| Sample | Dose | Intercept | slope | Volume weighted average domain size ($D_v$) (Å) (±10%) | Average microstrain ($\varepsilon$) ($10^{-3}$) (±5%) |
|---|---|---|---|---|---|
| Titanium | Unirradiated | 0.0013 | 0.0021 | 769 | 1.05 |
| | $1 \times 10^{17}\, O^{5+}/m^2$ | 0.0014 | 0.0066 | 714 | 3.30 |
| | $1 \times 10^{18}\, O^{5+}/m^2$ | 0.0014 | 0.0067 | 714 | 3.35 |
| | $1 \times 10^{19}\, O^{5+}/m^2$ | 0.0017 | 0.0061 | 588 | 3.05 |
| Ti-5%Ta-2%Nb | Unirradiated | 0.0016 | 0.0072 | 625 | 3.60 |
| | $1 \times 10^{17}\, O^{5+}/m^2$ | 0.0017 | 0.0070 | 588 | 3.50 |
| | $1 \times 10^{18}\, O^{5+}/m^2$ | 0.002 | 0.0058 | 500 | 2.90 |
| | $1 \times 10^{19}\, O^{5+}/m^2$ | 0.0024 | 0.0051 | 417 | 2.55 |

**Table-3 : Microstrain, dislocation density and stacking fault probabilities for Ti and Ti-5%Ta-2%Nb at different doses**

| Samples | Dose → | Micro strain ($10^{-3}$) Max. error ±0.00005 | | | | Dislocation density ($10^{15}$)($m^{-2}$) Max. error ±($6\times10^{14}$) | | | | Stacking faults at dose $\alpha$ ($10^{-3}$) Max. error | | |
|---|---|---|---|---|---|---|---|---|---|---|---|---|
| | | Unirr. | $10^{17}$ $O^{5+}/m^2$ | $10^{18}$ $O^{5+}/m^2$ | $10^{19}$ $O^{5+}/m^2$ | Unirr. | $10^{17}$ $O^{5+}/m^2$ | $10^{18}$ $O^{5+}/m^2$ | $10^{19}$ $O^{5+}/m^2$ | | | |
| Titanium | Fault un-affected | | | | | | | | | | | |
| | 002 | 1.3 | 2.2 | 2.6 | 2.5 | 0.5 | 2.0 | 2.3 | 2.2 | | | |
| | 004 | 1.3 | 2.2 | 2.6 | 2.5 | 0.5 | 2.0 | 2.3 | 2.2 | ±0.002x$10^{-3}$ | | |
| | 100 | 1.4 | 2.0 | 2.7 | 1.8 | 0.6 | 1.7 | 2.4 | 1.7 | | | |
| | 110 | 1.0 | 2.1 | 1.9 | 1.5 | 0.4 | 1.8 | 1.9 | 1.4 | | | |
| | 112 | 1.2 | 2.2 | 2.3 | 2.2 | 0.5 | 1.9 | 2.1 | 2.0 | $10^{17}$ $O^{5+}/m^2$ | $10^{18}$ $O^{5+}/m^2$ | $10^{19}$ $O^{5+}/m^2$ |
| | **Fault affected** | | | | | | | | | | | |
| | **101** | 1.3 | 2.2 | 2.6 | 2.1 | 0.5 | 1.8 | 2.3 | 2.0 | | | |
| | **202** | 1.3 | 2.2 | 2.7 | 2.1 | 0.5 | 1.8 | 2.3 | 2.0 | -0.04 | -0.05 | 0.03 |
| | **102** | 1.3 | 2.2 | 2.7 | 2.3 | 0.4 | 1.8 | 2.3 | 2.1 | | | |
| | **103** | 1.3 | 2.3 | 2.7 | 2.4 | 0.4 | 1.9 | 2.3 | 2.1 | | | |
| | **104** | 1.3 | 2.6 | 2.6 | 2.4 | 0.4 | 2.2 | 2.3 | 2.2 | | | |
| Ti-5%Ta-2%Nb | Fault un-affected | | | | | | | | | | | |
| | 002 | | 2.6 | 1.9 | 2.0 | 0.5 | 2.6 | 2.4 | 2.6 | | | |
| | 004 | | 2.6 | 1.9 | 2.0 | 0.5 | 2.6 | 2.4 | 2.6 | ±0.005x$10^{-3}$ | | |
| | 100 | | 4.7 | 1.8 | 1.7 | 0.4 | 4.8 | 2.3 | 2.6 | | | |
| | 110 | | 2.7 | 1.8 | 1.6 | 0.6 | 2.8 | 2.3 | 2.4 | | | |
| | 112 | | 2.9 | 1.8 | 1.9 | 0.8 | 2.9 | 2.4 | 2.0 | $10^{17}$ $O^{5+}/m^2$ | $10^{18}$ $O^{5+}/m^2$ | $10^{19}$ $O^{5+}/m^2$ |
| | **Fault affected** | | | | | | | | | | | |
| | **101** | | 3.5 | 1.8 | 1.9 | 0.6 | 3.4 | 2.3 | 3.2 | 1.4 | -1.7 | 1.6 |
| | **202** | | 3.5 | 1.8 | 1.9 | 0.6 | 3.4 | 2.3 | 3.2 | | | |
| | **102** | | 3.2 | 1.9 | 2.0 | 0.6 | 3.0 | 2.3 | 3.2 | | | |
| | **103** | | 3.1 | 1.9 | 2.0 | 0.6 | 2.9 | 2.3 | 3.2 | | | |
| | **104** | | 3.0 | 1.9 | 2.0 | 0.6 | 2.8 | 2.3 | 3.1 | | | |

**Table-4 : Results of Double Voigt Method for Ti and Ti-5%Ta-2%Nb at different doses**

| Samples | Dose | $\beta_{SC}$ $(10^{-2})$ | $\beta_{SG}$ $(10^{-2})$ | $\beta_{DC}$ $(10^{-2})$ | $\beta_{DG}$ $(10^{-2})$ | $D_S$ (Å) | $\varepsilon$ $(10^{-3})$ | $D_V$ (Å) |
|---|---|---|---|---|---|---|---|---|
| Titanium [001] | Unirradiated | 0.06 | 0.16 | 0.32 | 0 | 419±36 | 0.68±0.04 | 465±38 |
| | $1 \times 10^{17}$ $O^{5+}/m^2$ | 0.16 | 0.16 | 0.11 | 0.16 | 299±31 | 2.60±0.09 | 350±30 |
| | $1 \times 10^{18}$ $O^{5+}/m^2$ | 0.26 | 0.10 | 0.02 | 0.23 | 190±23 | 2.50±0.09 | 320±32 |
| | $1 \times 10^{19}$ $O^{5+}/m^2$ | 0.20 | 0.18 | 0.05 | 0.19 | 243±21 | 2.40±0.08 | 297±30 |
| Ti-5%Ta-2% Nb [001] | Unirradiated | 0.30 | 0 | 0.02 | 0.35 | 163±12 | 3.51±0.05 | 326±23 |
| | $1 \times 10^{17}$ $O^{5+}/m^2$ | 0.31 | 0 | 0.003 | 0.25 | 157±15 | 2.40±0.04 | 315±26 |
| | $1 \times 10^{18}$ $O^{5+}/m^2$ | 0.29 | 0.01 | 0.03 | 0.21 | 170±22 | 2.50±0.06 | 340±21 |
| | $1 \times 10^{19}$ $O^{5+}/m^2$ | 0.17 | 0.23 | 0.15 | 0 | 235±19 | 2.40±0.06 | 273±29 |

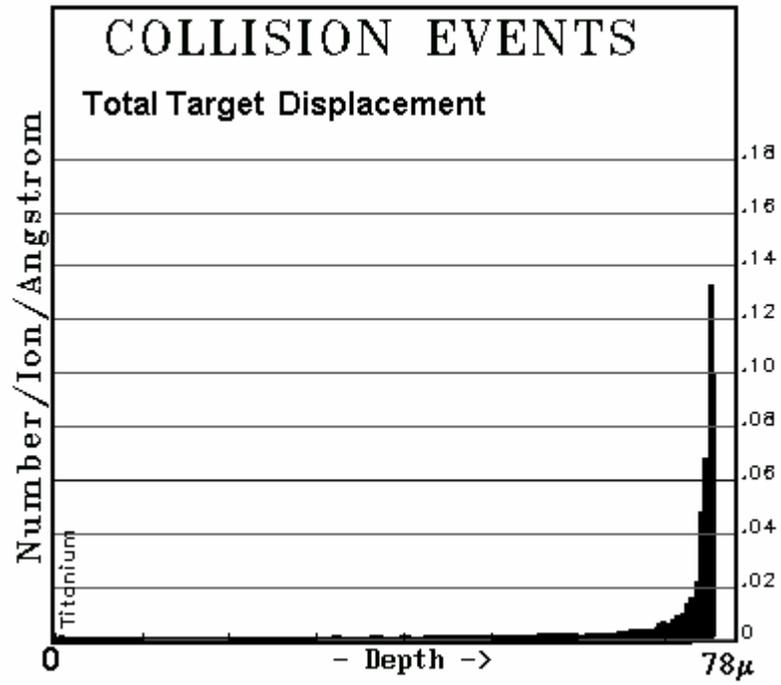

**Fig. 1. Damage profile of 116 MeV O$^{5+}$ in Titanium**

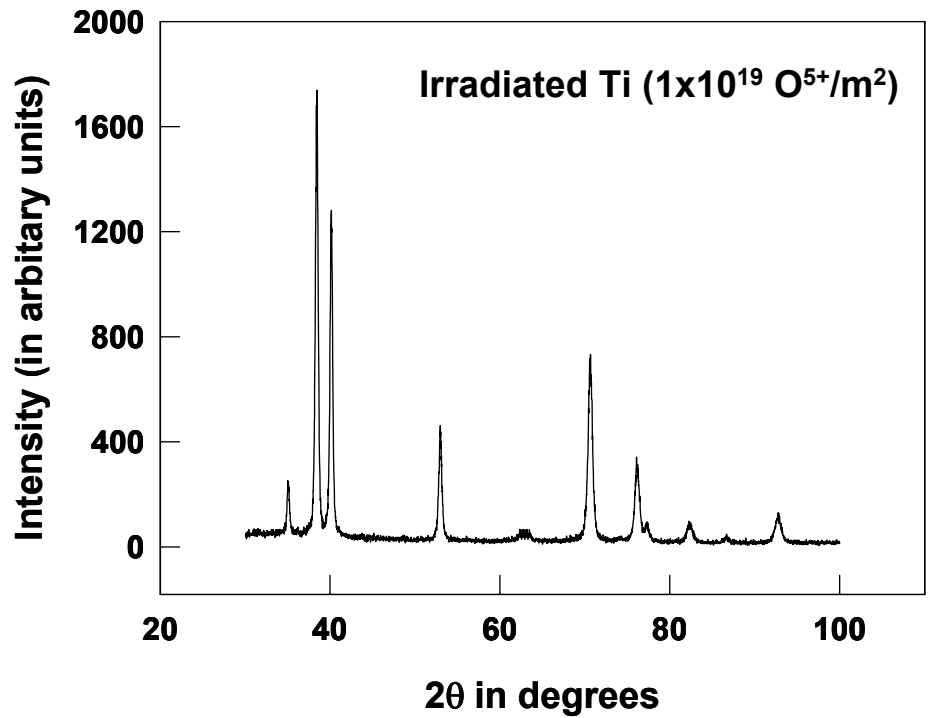

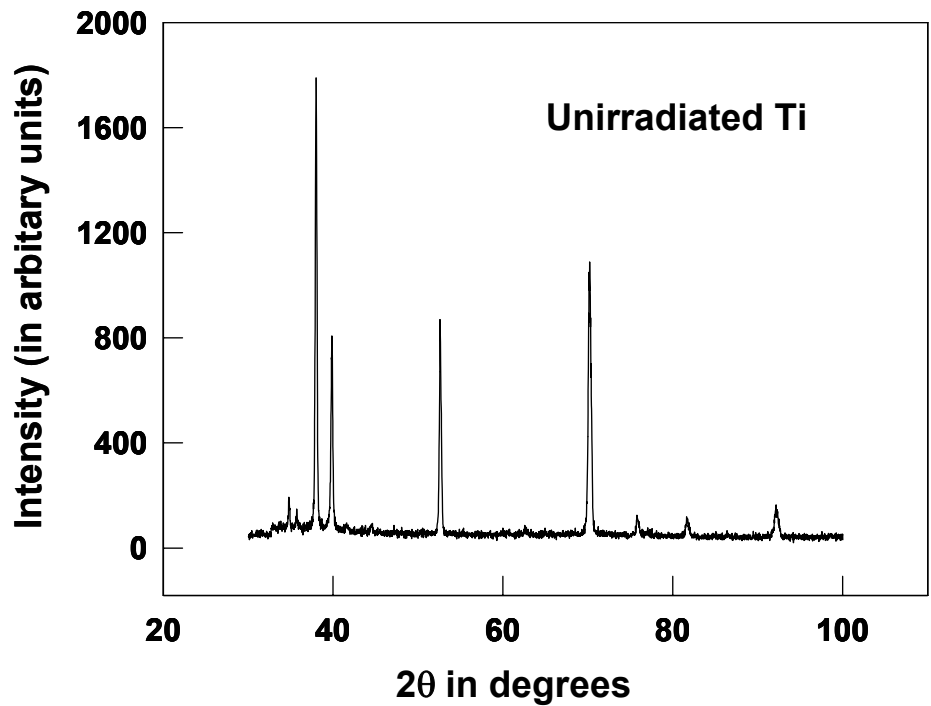

**Fig. 2. Typical XRD patterns of unirradiated and irradiated Ti**

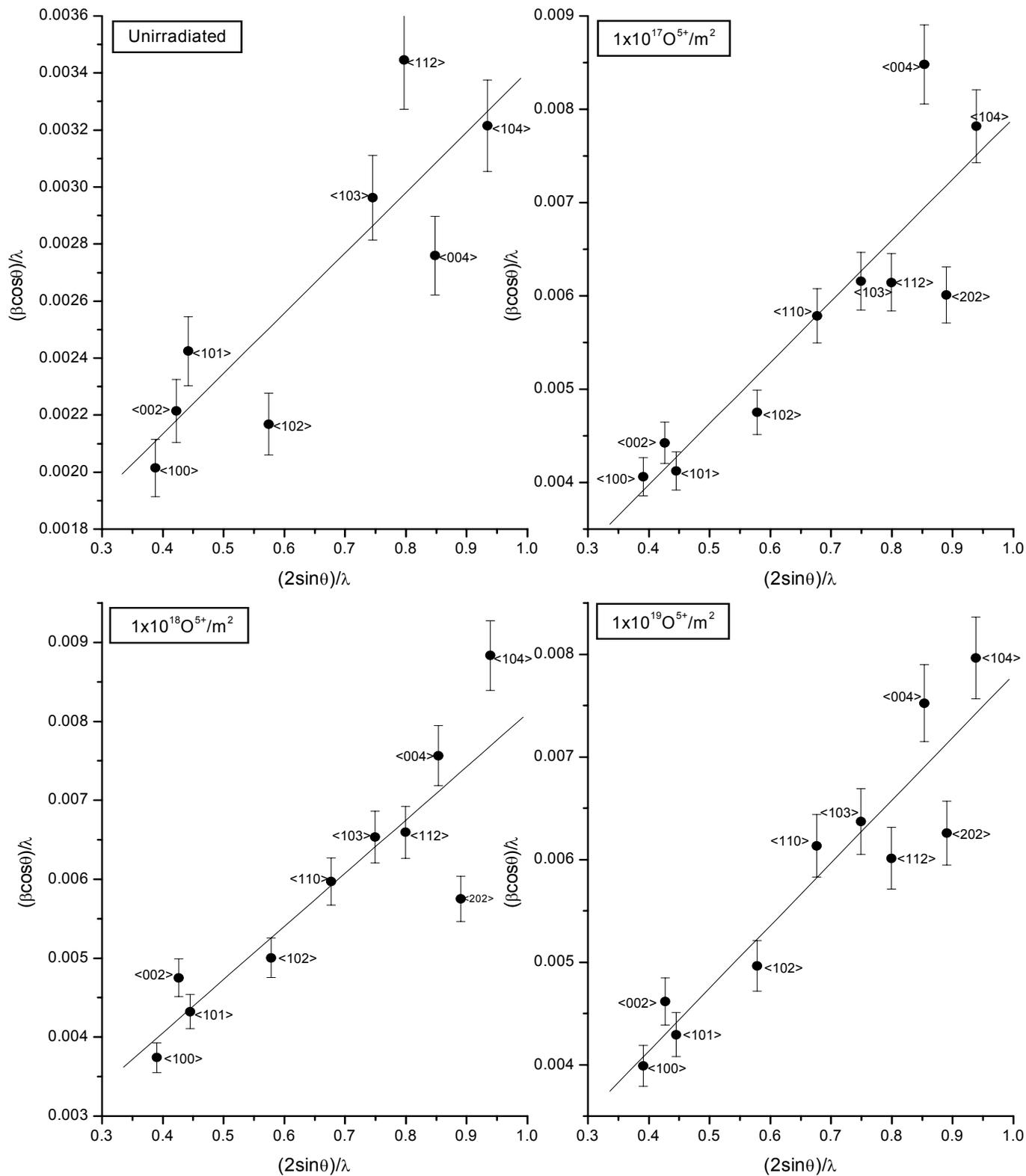

**Fig.3 Williamson-Hall Plots for unirradiated and irradiated Titanium at different doses**

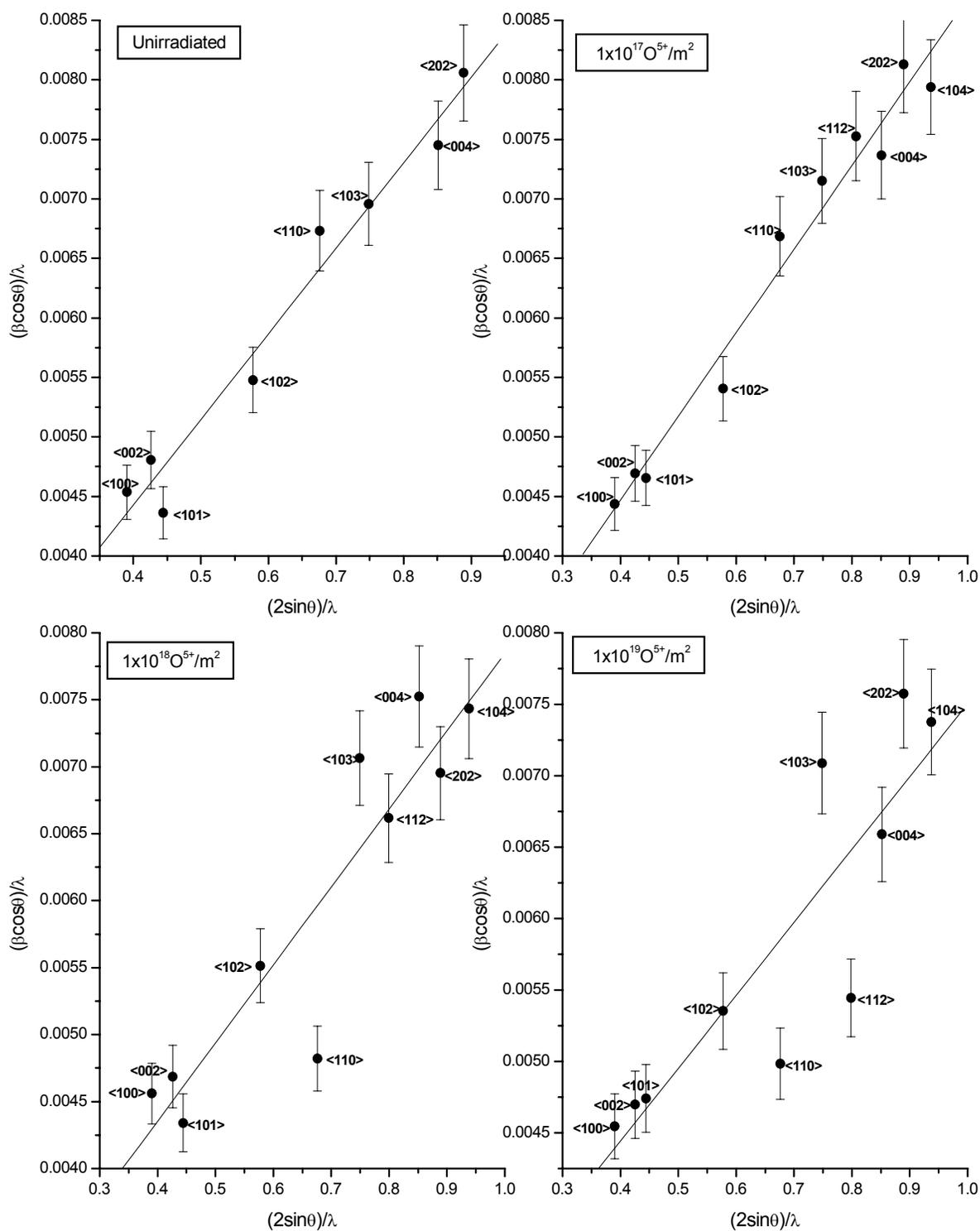

**Fig.4 Williamson-Hall Plots for unirradiated and irradiated Ti-5%Ta-2%Nb at different doses**

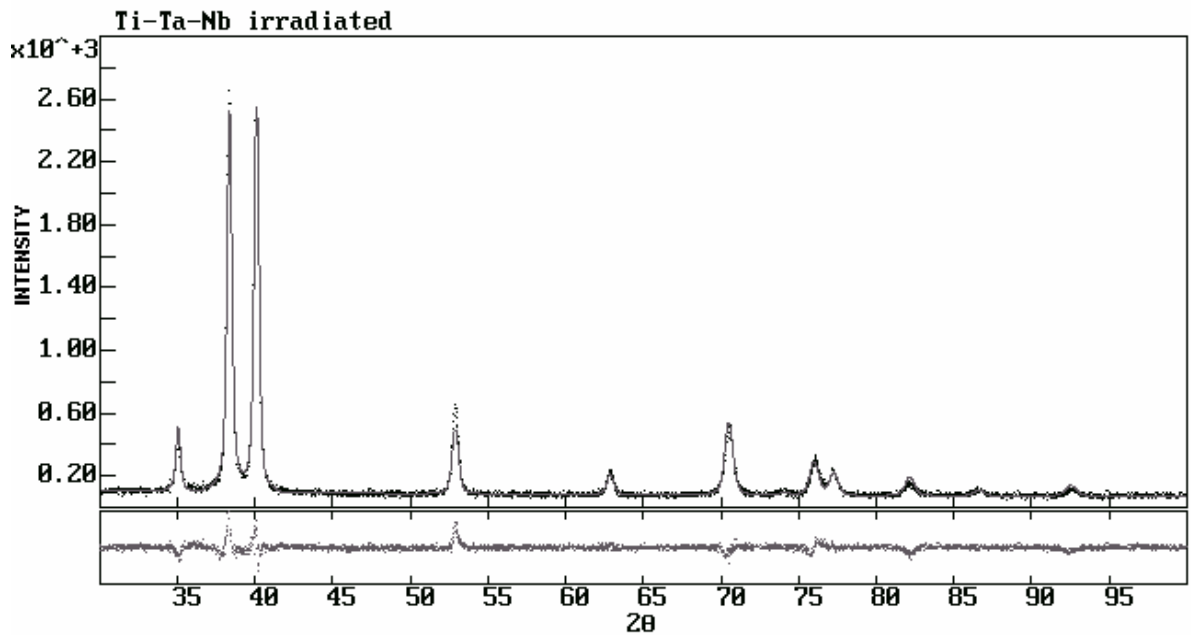

**Fig. 5. Rietveld fit for the diffraction profile of Ti-5%Ta-2%Nb, irradiated at $1 \times 10^{19} O^{5+}/m^2$. Residuals of the fit are also shown below.**

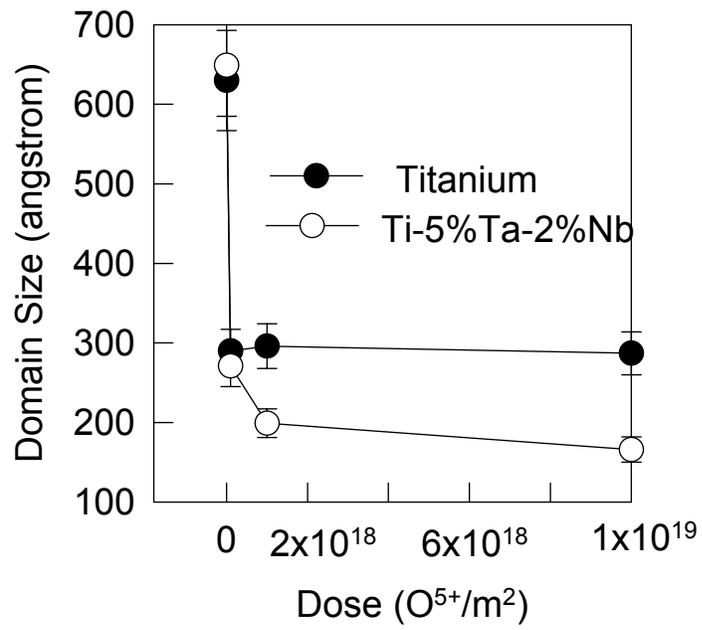

**Fig. 6. Variation of average domain size as a function of dose**

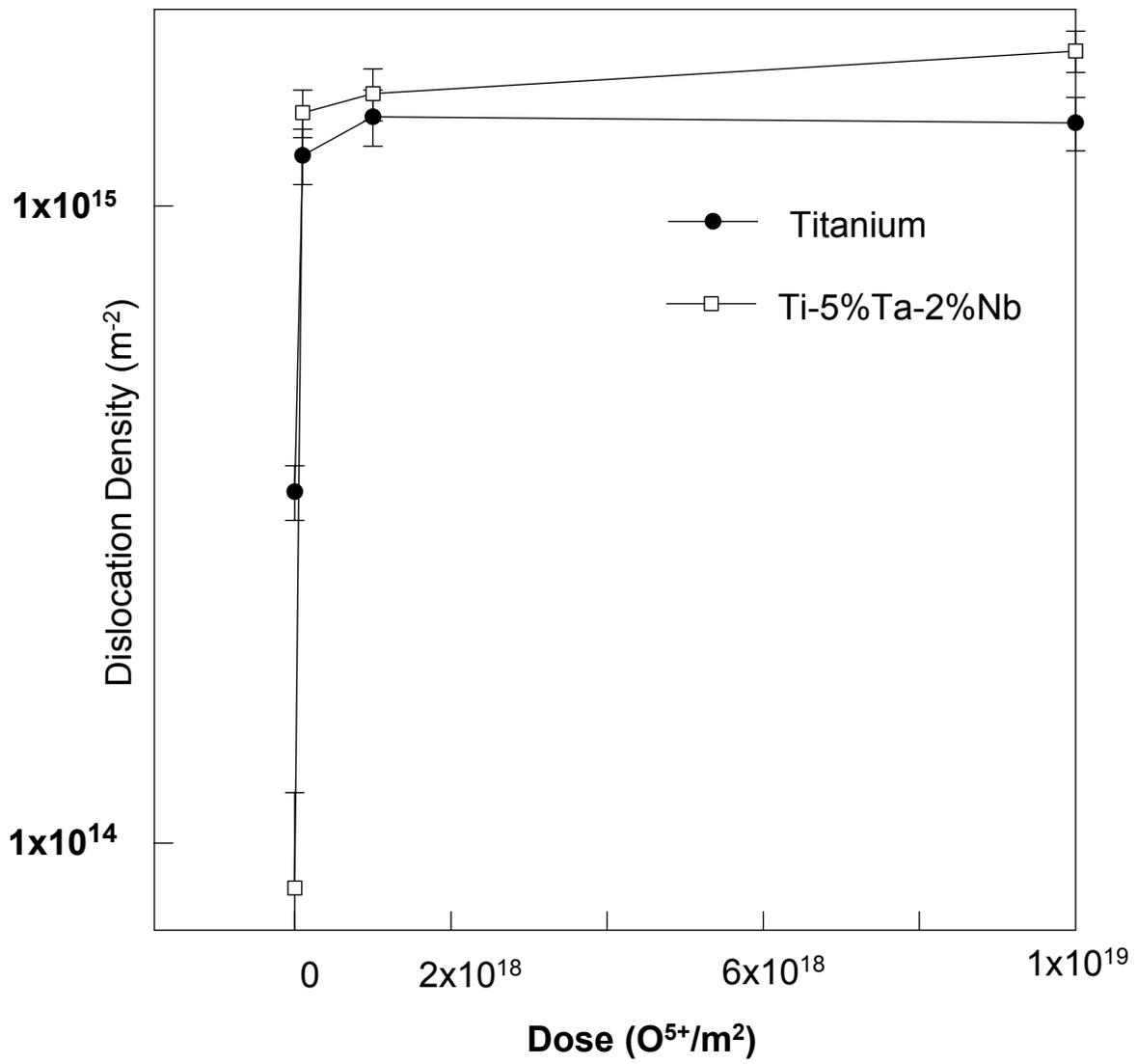

**Fig.7. Variation of Dislocation density as a function of dose**

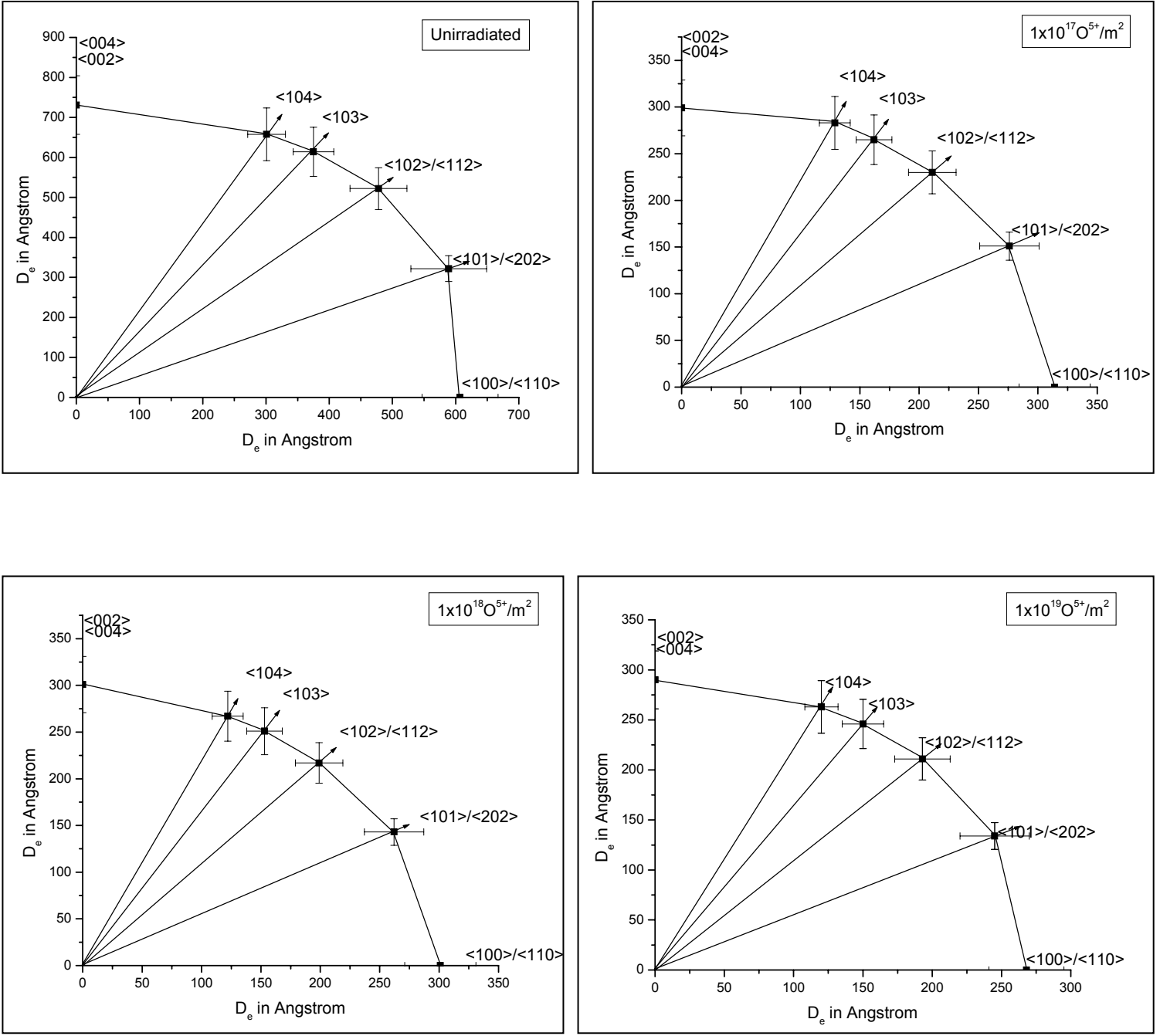

**Fig. 8. Projections of effective domain size on the plane containing the directions <002> and <100> (First quadrant) for unirradiated and irradiated Ti at different doses**

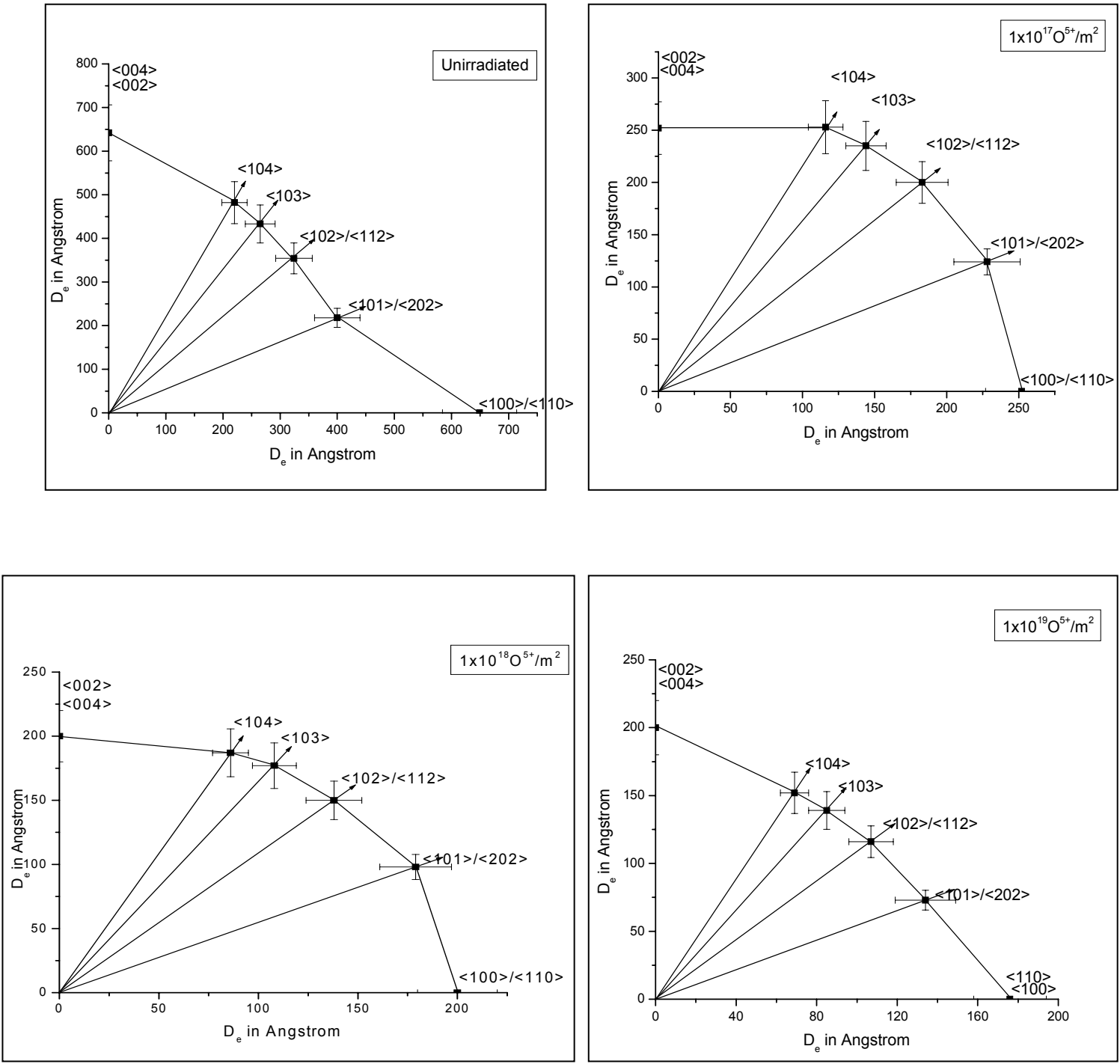

**Fig. 9. Projections of effective domain size on the plane containing the directions <002> and <100> (First quadrant) for unirradiated and irradiated Ti-5%Ta-2%Nb at different doses**

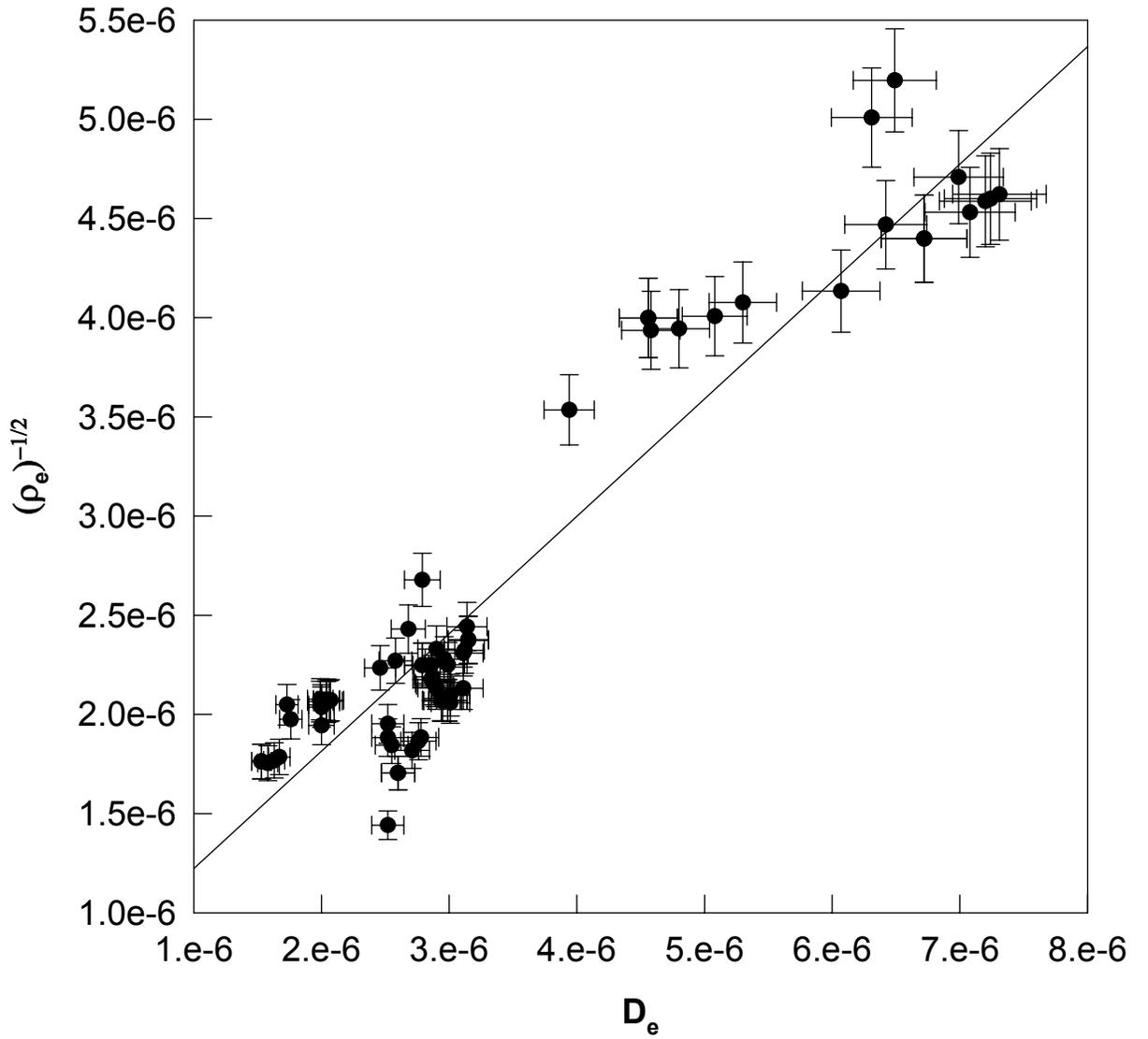

**Fig.10. Variation of $(\rho_e)^{-1/2}$ vs $D_e$**

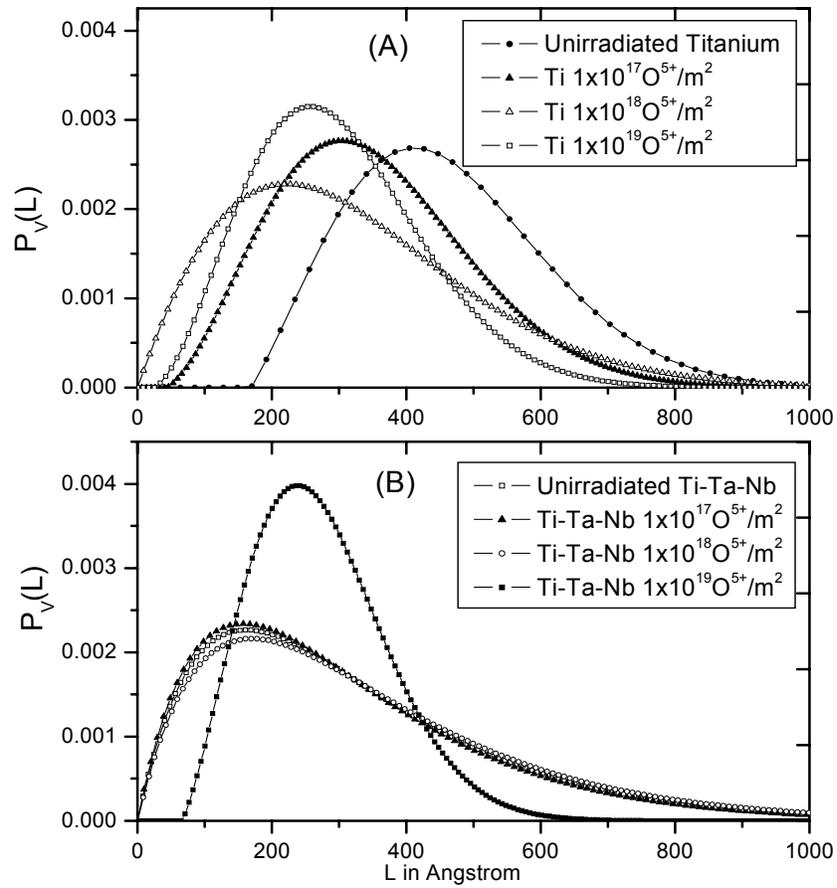

**Fig.11. Variation of domain size distibution $P_v(L)$ along <001>**
 **(A): Unirradiated and Irradiated Titanium**
 **(B): Unirradiated and Irradiated Ti-5%Ta-2%Nb**